\documentclass[lettersize,journal]{IEEEtran}
\usepackage{amsmath,amsfonts}
\usepackage{algorithmic}
\usepackage{algorithm}
\usepackage{array}
\usepackage[implicit=false]{hyperref}
\usepackage{tikz,xcolor}
\usepackage[caption=false,font=normalsize,labelfont=sf,textfont=sf]{subfig}
\usepackage{textcomp}
\usepackage{stfloats}
\usepackage{url}
\usepackage{verbatim}
\usepackage{graphicx}
\usepackage{cite}
\usepackage{amsmath}
\usepackage{booktabs}
\usepackage{multirow}
\usepackage{multicol}
\hypersetup{hidelinks,
	colorlinks=true,
	allcolors=black,
	pdfstartview=Fit,
	breaklinks=true}
\def\BibTeX{{\rm B\kern-.05em{\sc i\kern-.025em b}\kern-.08em
		T\kern-.1667em\lower.7ex\hbox{E}\kern-.125emX}}
\definecolor{lime}{HTML}{A6CE39}
\DeclareRobustCommand{\orcidicon}{%
	\begin{tikzpicture}
		\draw[lime, fill=lime] (0,0) 
		circle [radius=0.16] 
		node[white] {{\fontfamily{qag}\selectfont \tiny ID}};    \draw[white, fill=white] (-0.0625,0.095) 
		circle [radius=0.007];    \end{tikzpicture}
	\hspace{-2mm}}
\foreach \x in {A, ..., Z}{%
	\expandafter\xdef\csname orcid\x\endcsname{\noexpand\href{https://orcid.org/\csname orcidauthor\x\endcsname}{\noexpand\orcidicon}}
}

\begin{document}
	\begin{sloppypar}
\title{Deep Reinforcement Learning-aided Transmission Design for Energy-efficient Link Optimization \protect\\
	in Vehicular Communications\\
}

\author{Zhengpeng Wang\orcidA{},~\IEEEmembership{Student Member,~IEEE, }Yanqun Tang\orcidB{},~\IEEEmembership{Member,~IEEE,}\protect\\Yingzhe Mao\orcidC{},~\IEEEmembership{Student Member,~IEEE}, Tao Wang\orcidD{}, Xiunan Huang\orcidE{}
\thanks{Our earlier work on scenario identification has been accepted by the IEEE Wireless Communications and Networking Conference (WCNC), Dubai, Apr 2024\cite{wang2024}.}
	% <-this % stops a space
\thanks{This work was supported by Guangdong Natural Science Foundation under Grant 2019A1515011622. \emph{(Corresponding author: Yanqun Tang.)}}% <-this % stops a space
\thanks{Zhengpeng Wang, Yanqun Tang, Yingzhe Mao, Tao Wang and Xiunan Huang are with the School of Electronics and Communication Engineering, Sun Yat-sen University, China (email: wangzhp26@mail2.sysu.edu.cn; tangyq8@mail.sysu.edu.cn; maoyz@mail2.sysu.edu.cn; wangt369@mail2.sy\protect\\su.edu.cn; huangxn36@mail2.sysu.edu.cn).}}


\maketitle

\begin{abstract}
This letter presents a deep reinforcement learning (DRL) approach for transmission design to optimize the energy efficiency in vehicle-to-vehicle (V2V) communication links. Considering the dynamic environment of
vehicular communications, the optimization
problem is non-convex and mathematically difficult to solve. Hence, we propose scenario identification-based double and Dueling deep Q-Network (SI-D3QN), a DRL
algorithm integrating both double deep Q-Network and Dueling deep Q-Network, for the joint design of modulation and coding scheme (MCS) selection and power control. To be more specific, we employ SI techique to enhance link performance and assit the D3QN agent in refining its decision-making processes. The experiment results demonstrate that, across various optimization tasks, our proposed SI-D3QN agent outperforms the benchmark algorithms in terms of the valid actions and link performance metrics. Particularly, while ensuring significant improvement in energy efficiency, the agent facilitates a 29.6\% enhancement in the link throughput under the same energy consumption.
\end{abstract}

\begin{IEEEkeywords}
Reinforcement learning, transmission design, energy efficiency, vehicular communications, SI-D3QN.
\end{IEEEkeywords}

\section{Introduction}
\IEEEPARstart{V}{ehicle-to-Everything} (V2X) communication has evolved into a pivotal interconnectivity technology that enables vehicles to exchange information with any entity in their surroundings. In this context, vehicular communication with high-reliability and low-latency is considered one of the key application scenarios for Long Time Evolution (LTE), 5G, and even future 6G technologies. 

However, due to the high-speed movement of vehicles, the performance of vehicular communications is easily disrupted by the dynamic changes in the surrounding environment. To align with the design philosophy of 5G wireless communication systems\cite{zhao2020reinforcement}, which aims for high spectral efficiency and high energy efficiency, vehicle-to-vehicle (V2V) links require flexible transmission mechanisms to ensure efficient and stable communication. Furthermore, the demand for information applications and message sharing requires frequent access to vehicular network servers and the Internet in high congested scenarios\cite{guo2019fast}, imposing higher pressure on the energy consumption. Therefore, both link communication quality and long-term energy efficiency are crucial indicators for V2V communication systems.

Transmitted over the V2V channels characterized by time-varying multipath fading, the fixed modulation and coding scheme (MCS) makes it difficult to achieve long-term energy-saving transmission. Furthermore, without finely adjusting transimission power levels, the links are unable to effectively counteract signal attenuation and interference\cite{aznar2021simultaneous}, thereby leading to a degradation in communication quality. Therefore, under the goal of energy-efficient transmission in vehicular communications, MCS selection and power control become two key interrelated approaches.

As V2V networks become increasingly dense and complex, the conventional optimization methods that necessitate precise mathematical models and expert knowledge  struggle to cope with dynamic environments. However, with the rapid development of artificial intelligence and machine learning \cite{ghadimi2017reinforcement}, deep reinforcement learning (DRL) has been considered the optimal technological pathway for addressing complex decision-making and non-convex optimization problems. To deal with the challenge of  energy-efficient transmission in the field of underwater acoustic communication, the authors in \cite{jing2022adaptive} proposed an adaptive coding and modulation scheme based on double deep Q-network (DDQN) in order to maximize the long-term energy efficiency. In the literature \cite{parsa2022joint}, the authors introduced an intelligent energy-efficient link adaptation agent in 5G NR to find the best match between the channel condition and the link parameters. Experiments  indicate that the DRL algorithm significantly improves link efficiency and throughput. The authors \cite{zhang2021deep} proposed a solution to the power/rate control problem in multi-user V2V networks using the deep deterministic policy gradient (DDPG) algorithm, demostrating the advantages of DRL agents applied into wireless communication systems.

Most pioneer works in this field rely on basic Q-learning and its variants, which tend to overestimate the value of actions, leading to suboptimal decision-making. Furthermore, agents designed for rapidly changing environmental states often face issues akin to the cold-start problem\cite{ungar2002methods} in recommendation tasks. Additionally, as the potential state and action spaces become more complex, the experience accumulated by these basic reinforcement learning algorithms fails to generalize across similar states\cite{zhang2021deep}.

Inspired by state-of-the-art works, in this paper, we propose a novel DRL framework, which optimizes link performance through joint design of MCS selection and power control in vehicular communications. Major contributions and novelties of this letter are summarized as follows:

\begin{itemize}
	\item[$\bullet$]To effectively explore the high-dimensional state-action space and reduce the overestimation problem during the learning process, we propose a new combination of DDQN and Dueling deep Q-Network (DQN) algorithm, named D3QN. This structure not only leverages the strengths of Dueling DQN by learning representations of state-value and action advantage, but also delegates action selection and evaluation to two independent Q-Networks by incorporating the characteristics of DDQN.
	\item[$\bullet$] We innovatively integrate scenario identification (SI) technology to the D3QN agent's state space design, which is further referred to as SI-D3QN. This step significantly accelerates the agent's understanding of different vehicular scenarios and adaptation to unknown environments, allowing it to adjust strategies even in highly dynamic vehicular environment.
	\item[$\bullet$]We consider two types of link optimization problems, modeling them as action-reward non-entangled and action-reward entangled forms. Through the effective interaction between the SI-D3QN agent and the environment, the V2V link does not sacrifice communication quality for higher  energy efficiency, and may even see anticipated improvements in complex entangled conditions.
	\item[$\bullet$]We conduct a detailed comparison of the proposed SI-D3QN agent with DRL benchmaker algorithms, particle swarm optimization (PSO), simulated annealing (SA), fixed transmission scheme, and random decision strategy. Extensive experiments demonstrate that, even in the rapidly changing vehicular environments, the SI-D3QN agent exhibits superior performance in terms of valid decisions and long-term link performance, fully showcasing the significant advantages of the SI-D3QN agent in vehicular transmission design.
\end{itemize}
\section{System Description and Problem Formulation}
In this section, we will introduce our system model and present the formulation of the energy efficiency optimization problem in vehicular communications. 
\subsection{System Model}\label{AA}
We consider a typical V2V system, comprised of a transmitter-receiver pair. The system adheres to IEEE 802.11p, which is the physical layer standard especially for dedicated short-range communication \cite{zhu2018big}. According to the standard, it supports eight types of MCS, including BPSK 1/2, BPSK 3/4, QPSK 1/2, QPSK 3/4, 16QAM 1/2, 16QAM 3/4, 64QAM 2/3 and 64QAM 3/4.

Let $\mathcal{R}\!=\!\{R_{\mathrm{1}},R_{\mathrm{2}},\cdots,R_{J}\}$ and $\mathcal{M}\!=\!\{M_{\mathrm{1}},M_{\mathrm{2}},\cdots,M_{K}\}$ are defined as the finite set of discrete code rates and modulation sizes. Thus, $\mathcal{Q}\!=\!\{Q_{(R_{\mathrm{1}},M_{\mathrm{1}})},Q_{(R_{\mathrm{1}},M_{\mathrm{2}})},\ldots,Q_{(R,M)}\}$ represents the set of supported MCS, where $R\in \mathcal{R}$ and $M \in \mathcal{M}$ denote the selected code rate and modulation scheme. The transmitter first uses a convolutional channel coder with code rate $R$ and then the coded bits are further converted to symbols in complex by constellation mapping through modulation scheme $M$. Next, the orthogonal frequency division multiplexing (OFDM) modulation is realized by using 64-point inverse fast Fourier transform (IFFT) and the cyclic prefixs (CPs) are inserted for eliminating inter-symbol interference (ISI). Afterwards, the OFDM symbols are transmitted with a certain transmission power $P$ through the V2V channel with both time-varying and fading characteristics and then get distorted to the receiver of the vehicular link, where $P \in\mathcal{P}$, $\mathcal{P}=\{P_{1},P_{2},\ldots,P_{L}\}$. The receiver processes the received signal in a reverse manner with the help of equalizers and estimators and finally performs demodulation and decoding operations.

Furthermore, according to relevant study\cite{yang2022mobilenet}, there are five typical scenarios considered in vehicular communications, namely rural LOS (R-LOS), urban approaching LOS (U-A-LOS), urban NLOS (U-NLOS), highway LOS (H-LOS), and highway NLOS (H-NLOS). Every scenario contains its corresponding channel characteristic, including avarage path gain, path delay and Doppler shift. Each multipath tap $m= 1, 2, ... , M $ of the time-domain channel impulse response for the V2V channel can be modeled as 
\begin{equation}
h(t,\tau)=\sum_{m=1}^MA_me^{j2\pi\upsilon_mt}\delta(\tau-\tau_m),\end{equation} where $A_m$,$\tau_m$,$\upsilon_m$ represent the amplitude, delay and Doppler frequency respectively.
Hence, in our system, the receiver utilizes scenario identification techique \cite{wang2024} to extract the characteristic parameters of different vehicular scenarios and estimate the channel responses from the long training symbols (LTS) in the OFDM frame. 
Here, the received frequency response on the subcarrier $n$ of the two identical LTS are given as follows.
\begin{equation}
	Y^n_{\mathrm{1}}=H^n(f,\tau,A_m,\tau_m,\upsilon_m)X^n(f,\tau)+W^n_1,
\end{equation} 
\begin{equation}
	\begin{aligned}
		Y^n_{\mathrm{2}}=H^n(f,\tau,A_m,\tau_m,\upsilon_m)X^n(f,\tau)e^{j2\pi f\Delta {t}}+W^n_2,
	\end{aligned}
\end{equation} 
where $f\!$ and $\Delta t$  are the carrier frequency and time slot between the LTS respectively, $H^n(f,\tau,A_m,\tau_m,\upsilon_m)\!$ represents the frequency-domain channel impulse response, $W^n_1$ and $W^n_2$ are the complex additive white Gaussian noise and $X^n(f,\tau)$ is denoted as the frequency response of the LTS. The estimated SNR $\hat{\delta}$ can be expressed as
\begin{equation}
	\hat{\delta}=10\log_{10}\left(\frac {2N_{\mathrm{sc}}P_\mathrm{t}}{|\mathbf{Y}_{\mathrm{1}}-\mathbf{Y}_{\mathrm{2}}|^2}-1\right),
\end{equation}
where $P_\mathrm{t}$ represents the total signal power,\begin{figure}[ht]
	\hspace{1mm}\includegraphics[width=1\columnwidth,height=0.55\linewidth]{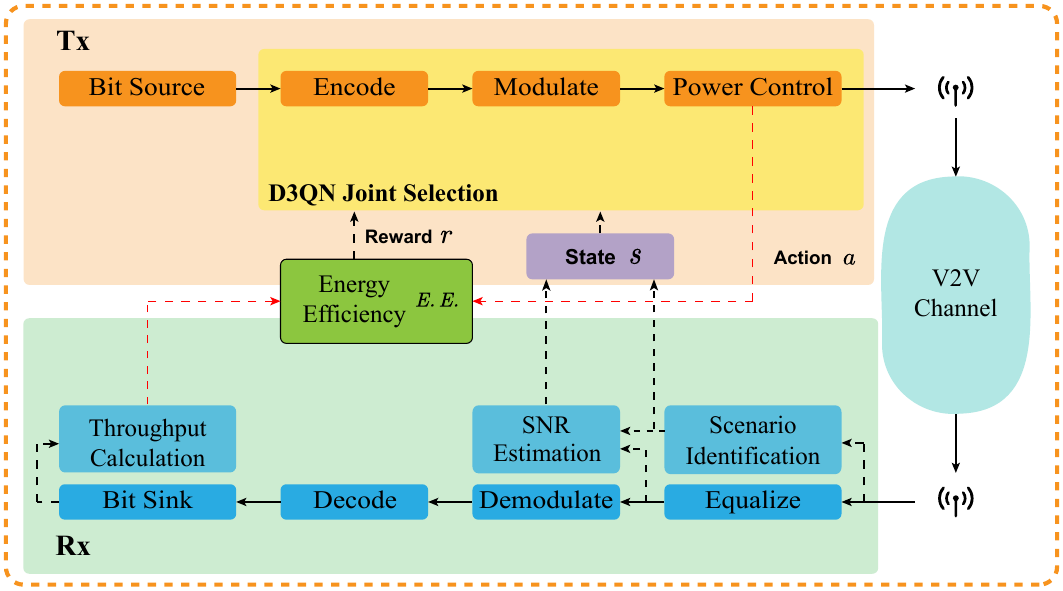}
	
	\caption{The structure of the D3QN agent with evaluation Dueling Q-Network and target Dueling Q-Network. }
	\label{fig4:env}
\end{figure} $N_{\mathrm{sc}} $ is the total number of the subcarriers, $\mathbf{Y}_{\mathrm{1}}$ and $\mathbf{Y}_{\mathrm{2}}$ represent the frequency response matrix of the LTS, 
as shown in Equation (5) and Equation (6):
\begin{equation}\left.\mathbf{Y}_{\mathrm{1}}=\left\{\begin{matrix}Y_{\mathrm{1}}^1,Y_{\mathrm{1}}^2,\ldots,Y_{\mathrm{1}}^{N_{\mathrm{sc}}}\end{matrix}\right.\right\},\end{equation} 

\begin{equation}\left.\mathbf{Y}_{\mathrm{2}}=\left\{\begin{matrix}Y_{\mathrm{2}}^1,Y_{\mathrm{2}}^2,\ldots,Y_{\mathrm{2}}^{N_{\mathrm{sc}}}\end{matrix}\right.\right\}.\end{equation} 

With the application of scenario identification, we can obtain the refined SNR estimation and improve link performance. The PHY throughput $T_{\mathrm{p}}$ of the data link is computed as	
\begin{equation}
	T_{\mathrm{p}}=\frac{n_{\mathrm{s}}N_{\mathrm{b}}}{n_{\mathrm{s}}t_{\mathrm{s}}+t_{\mathrm{o}}}\times(1-\mathrm{PER}),
\end{equation}where $n_{\mathrm{s}}$ is the number of data symbols in the packet, $\mathrm{PER}$ denotes the packet error rate (PER), $N_{\mathrm{b}}$ stands for the number of data bits carried by every symbol, $t_{\mathrm{s}}$ and $t_{\mathrm{o}}$ correspond to the symbol time and overhead time (including the training preamble symbol and the signal symbol) respectively.

The energy efficiency $\mathrm{EE}$ of the link is defined as communication quality per unit of consumed power maintained, which is calculated as follow:
 \begin{equation}
 \mathrm{EE}=\frac{T_{\mathrm{p}}}{P}.
 \end{equation}
 \subsection{Problem Formulation}
The target of transmission design is to jointly power control and MCS selection to optimize the energy efficiency of the vehicular link. The optimization problem can be mathematically formulated as:
\begin{equation}
	\begin{split}
	    	&\max_{\begin{array}{c}Q_{(R_n,M_n)},\forall n\\P_{n},\forall n\end{array}} \,\, \sum_{n=1}^{N}{\mathrm{EE}[n]}\\
		    &s.t.\quad  \left\{\begin{array}{lc}
		Q_{(R_n,M_n)}\in\mathcal{Q}, \forall n,\\
			P_{n}\in\mathcal{P}, \forall n,\\
			0\leq \mathrm{PER}[n]\leq \mathrm{PER}_{\mathrm{r}}, \forall n,
		\end{array}\right.
	\end{split}
\end{equation}
in which $\mathrm{EE}[n]$ is defined as the energy efficiency and $Q_{(R_n,M_n)}$, $P_{n}$ are the MCS and power chosen from the set of supported transmission scheme in the $n$-th transmission. Besides, in order to ensure the basic quality of communication, the PER in the link is constraint to the rated value $\mathrm{PER}_{\mathrm{r}}$.
Considering that the non-convex optimization problem is challenging and the performance indicators are highly entangled, in this letter, we employ the advanced DRL algorithm to find feasible solutions for this decision problem rather than solve it mathematically. 
\begin{algorithm}[ht]
	\caption{Double \& Dueling Deep Q-Network algorithm}\label{alg:alg1}
	\begin{algorithmic}[1]
		\STATE \textbf{Input:} Initialize evaluation Dueling Q-Network parameters $\theta$ and target Dueling Q-Network parameters $\theta^{-}$; \\ %算法的输入参数：Input
		\STATE \textbf{Input:} Initialize experience replay buffer $\mathcal{B}$ and initial exploration value of $\epsilon$-greedy policy; \\ %算法的输入参数：Input
		\STATE \textbf{for} every eposide $t=1,2,\ldots,T$ \textbf{do}
		\STATE \hspace{0.5cm} Generate initial state $\mathbf{s}_{t,1}$ for the first transmission
		\STATE \hspace{0.5cm} \textbf{for} every transmission $n=1,2,\ldots,N$ \textbf{do}
		
		\STATE \hspace{1cm}With probability $\epsilon_{t,n}$ to pick a random action $a_{t,n}\in\mathcal{Q}$, otherwise select an  action through Eq.(6).
		\STATE \hspace{1cm} Receive next state $\mathbf{s^{\prime}}_{t,n+1}$ and reward $r_{t,n}$ 
		\STATE \hspace{1cm} Store $(\mathbf{s}_{t,n},a_{t,n},r_{t,n},\mathbf{s^{\prime}}_{t,n+1})$ to replay buffer $\mathcal{B}$
		\STATE \hspace{1cm} \textbf{Train evaluation network every 1 step:}
		\STATE \hspace{1.2cm} Sample mini-batch of transitions from set $\mathcal{B}$
		\STATE \hspace{1.2cm} Train the Q network with the calculated loss $L(\theta)$
		\STATE \hspace{1.2cm} Update evaluation network parameters $\theta$
		\STATE \hspace{1cm} \textbf{Update target network every $N^{-}$ steps:}
		\STATE \hspace{1.2cm} Update $\theta^{-}\leftarrow\theta$
		\STATE \hspace{0.5cm}$ \textbf{end for} $
		\STATE $ \textbf{end for} $
	\end{algorithmic}
	\label{alg1}
\end{algorithm} 
\section{Deep Reinforcement Learning-aided Transmission Design}
The principle structure of SI-D3QN deployed in the vehicular communication system is shown in Fig. 1. Assuming that the feedback of the transmission over the link is instantaneous, the SI-D3QN agent jointly optimizes both MCS selection and power control to maximize the objective function in Eq. 9, by utilizing the type of vehicular scenario given by scenario identification \cite{wang2024} and the SNR estimated by the receiver as input state information.

\subsection{Markov Decision Process Modeling}
Reinforcement Learning (RL) is a branch of machine learning field that enables an agent to learn through continuous interaction with the environment, aiming to optimize its performance. Usually, a RL problem is modeled as a Markov Decision Process (MDP), which provides a mathematical framework to deal with optimal sequences of actions\cite{aznar2021simultaneous}. In order to obtain the optimal policy, we derive the agent, states, actions, rewards and transition probability through a MDP framework in vehicular communications. In the following, we introduce them one by one in detail.

\begin{enumerate}
	\item \textbf{\emph{Agent}}: The agent, designated as SI-D3QN, is deployed in the transmitter component of the V2V pair.
	\item \textbf{\emph{States}}: The state $\mathbf{s}\in\mathcal{S}$ is defined as a vector that involves the channel conditions and the properties of the agent, with $\mathcal{S}$ being the infinite set of possible states. For each V2V link during the \(n\)-th transmission, the observed state, denoted as \(\mathbf{s}\), comprises triplets formulated as \(\mathbf{s} = \{\psi_n, \delta_n, n\}\). Here, \(\psi_n\) characterizes the current type of vehicular scenario.
	\item \textbf{\emph{Actions}}: The action $a\in\mathcal{A}$ is to jointly determine the MCS selection and transmit power from the available action set $\mathcal{A}$, which is defined as $\mathcal{A}=\mathcal{Q} \times\mathcal{P}$.
	\begin{figure}[ht]
		\hspace{1mm}\includegraphics[width=1\columnwidth,height=0.55\linewidth]{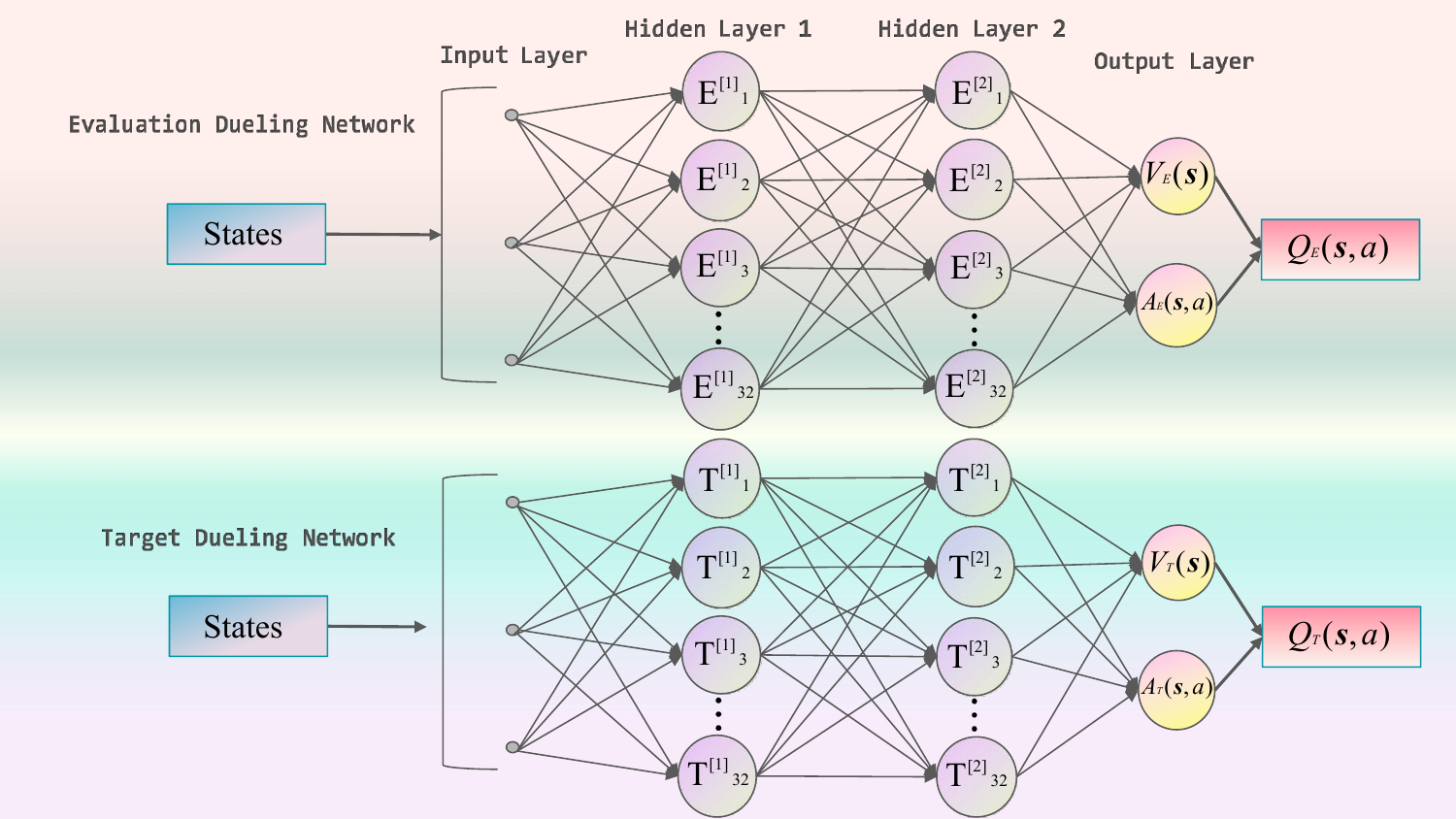}
		
		\caption{The structure of the SI-D3QN agent with evaluation Dueling Q-Network and target Dueling Q-Network. }
		\label{fig4:env}
	\end{figure}
	\item \textbf{\emph{Rewards}}: The reward $\!r\!$ is the energy efficiency $\!EE[n]\!$ in the $n$-th transmission, with its magnitude reflecting the effectiveness of the actions taken by the agent. Our goal is to maximize the cumulative (long-term) reward $\sum_{n=1}^{N}{EE[n]}$ after $N$ transmissions and make sure high quality of vehicular communications.
	\item \textbf{\emph{Transition probabilities:}} We simulate the complex vehicular communication environment with a randomly varying SNR $\gamma_n$. Consequently, the transition from one state $\mathbf{s}$ to any subsequent state $\mathbf{s}^{\prime}$ is entirely stochastic, continuing until the end of $N$ transmissions.
\end{enumerate}

\subsection{SI-D3QN: SI-Based Double \& Dueling Deep Q-Network }
The structure of the SI-D3QN agent is shown in Fig. 2, which consists of two full connected layers with 32 neurons each. As illustrated in Algorithm 1, it provides a detailed description of the principles of D3QN algorithm.

The training of the SI-D3QN agent consists of four main steps: Firstly, build up two deep Q-Networks with the Dueling architecture,  which has two separate estimators: one estimator is used to estimate the state-value function $V(\mathbf{s})$ and the other provides a reasonable estimate of the advantage function $A(\mathbf{s},a)$. The action-value function is calculated by these two estimators as following  
\begin{equation}
Q(\mathbf{s},a;\theta,w_{\alpha,\beta})=V(\mathbf{s};\theta,\beta)+A(\mathbf{s},a;\theta,\alpha).
\end{equation} 
Here, $\alpha$ and $\beta$ are the parameters of the two streams of fully-connected layers\cite{wang2016dueling}, while $w_{\alpha,\beta}$ is denoted as the hybrid weights of the two streams. Secondly, the two Dueling networks, named the evaluation Dueling network and the target Dueling network, are decoupled for selection and evaluation. The policy $\pi(\mathbf{s})$ for decision making can be expressed as
\begin{equation}
	\pi(\boldsymbol{s})=\left\{\begin{array}{c}\text{a random action in }\mathcal{A},\text{with prob. }\epsilon,\\\mathop{\arg \max}\limits_{a\in\mathcal{A}}Q(\boldsymbol{s},a;\theta,w_{\alpha,\beta}),\text{with prob. }1-\epsilon,\end{array}\right.
\end{equation} 
where $\epsilon$ is the exploration rate of greedy policy.
The action-value function $Q(\mathbf{s},a;\theta,w_{\alpha,\beta})$  is updated as
\begin{table}[ht] 
	\setlength{\arrayrulewidth}{0.3mm}
	\setlength{\tabcolsep}{11.3pt}
	\renewcommand{\arraystretch}{1.08}
	\label{tab:univ-compa}
	\caption{Key System Parameters.}
	\scalebox{1.05}
	{\begin{tabular}{c c}
			\hline
			\textbf{Parameters}& \textbf {Values}\\
			\hline
			Transmission scheme&OFDM\\
			Bandwidth (MHz)& 10\\
			Vehicular scenario& Urban NLOS\\
			Data rate (Mbps)&3, 4.5, 6, 9, 12, 18, 24, 27\\
			Modulation schemes& BPSK, QPSK, 16QAM, 64QAM\\			 Convolutional coding rate& 1/2, 2/3, 3/4\\
			Transmission power (W) &0.6, 0.8, 1, 1.2, 1.4\\
			Number of subcarriers&64\\
			FFT size&64\\
			\hline
	\end{tabular}}
\end{table}
\begin{table}[h]
	\centering
	\setlength{\arrayrulewidth}{0.35mm}
	\setlength{\tabcolsep}{1.5pt}
	\renewcommand{\arraystretch}{0.95}
	\label{tab:univ-compa}
	\caption{Simulation Parameters of Different Algorithms.}
	\scalebox{0.7}
	
	\begin{tabular}{c c|c c|c c}
		\hline
		DRL	Param & Val & PSO Param& Val& SA Param& Val \\
		\hline
		Batchsize & 32 & Particles number & 50 & Particles number & 50\\
		Discount factor & 0.99 & Inertia weight & 0.6& Inertia weight & 0.6 \\
		Learning rate & 0.01 & Learning factor 1 & 1.2 & Learning factor 1 & 1.2\\
		Replay buffer& 20000  & Learning factor 2 &1.8&  Learning factor 2 &1.8\\
		\(\epsilon \) & [0.01, 1]   & Velocity limits & None& Initial temperature & 450\\
		\(N^{-}\) & 1000 &Position limits & None  & Final temperature & 0\\
		\hline
	\end{tabular}
\end{table}
\begin{equation}
		\begin{aligned}
	Q(\mathbf{s},a;\theta,w_{\alpha,\beta})\leftarrow Q(\mathbf{s},a;\theta,w_{\alpha,\beta})+\\\kappa[R+\gamma\max_{a^{\prime}\in\mathcal{A}}Q(\mathbf{s}^{\prime},a^{\prime};\theta,w_{\alpha,\beta}&)-Q(\mathbf{s},a;\theta,w_{\alpha,\beta})],
		\end{aligned}
		\end{equation}
by minimizing the mean square error (MSE) loss function, which is given by
\begin{equation}
	\begin{aligned}
&L(\theta)=(R+\gamma\max_{a\in \mathcal{A}}Q(\mathbf{s}^{\prime},a;\theta^-,w_{\alpha,\beta}^-)-Q(\mathbf{s},a;\theta,w_{\alpha,\beta}))^2,
	\end{aligned}
\end{equation}
where $\gamma$ represents the discount factor, $R$ is the reward for action $a$, $\kappa$ is the learning rate, $a^{\prime}$ represents the action in the next state $\mathbf{s}^{\prime}$ and $w_{\alpha,\beta}^-$ is denoted as the weights of two streams for the target network. Thirdly, experience replay mechanism is another key concept of SI-D3QN. The experience of every step is stored in the replay buffer $\mathcal{B}$ and sampled for updating the evaluation network parameters $\theta$. Every $N^{-}$ training steps, the weights of target Dueling model will be transformed from $\theta$ to $\theta^{-}$. Finally, the policy $\pi(\mathbf{s})$  can converge towards optimality after sufficient iterations, at which point the solution to the original problem (9) can be obtained.

Built upon the foundations of two advanced DRL frameworks, the SI-D3QN agent is capable of more accurately choosing the optimal action in the current state during policy evaluation and effectively reduces overoptimistic value estimation of the actions.

\section{Simulation Results and Analysis}
In this section, we evaluate the performance of the proposed SI-D3QN agent in two representative scenes based on game
theory for V2V communication, which can be considered as action-reward non-entangled and action-reward entangled optimization. We treat them as two non-cooperative games to verify the robustness of SI-D3QN. The DRL experiments are executed on a NVIDIA RTX 4090 GPU and i9-13900K CPU. 

\begin{figure}[ht]
	\subfloat[]{\includegraphics[width=0.23\textwidth]{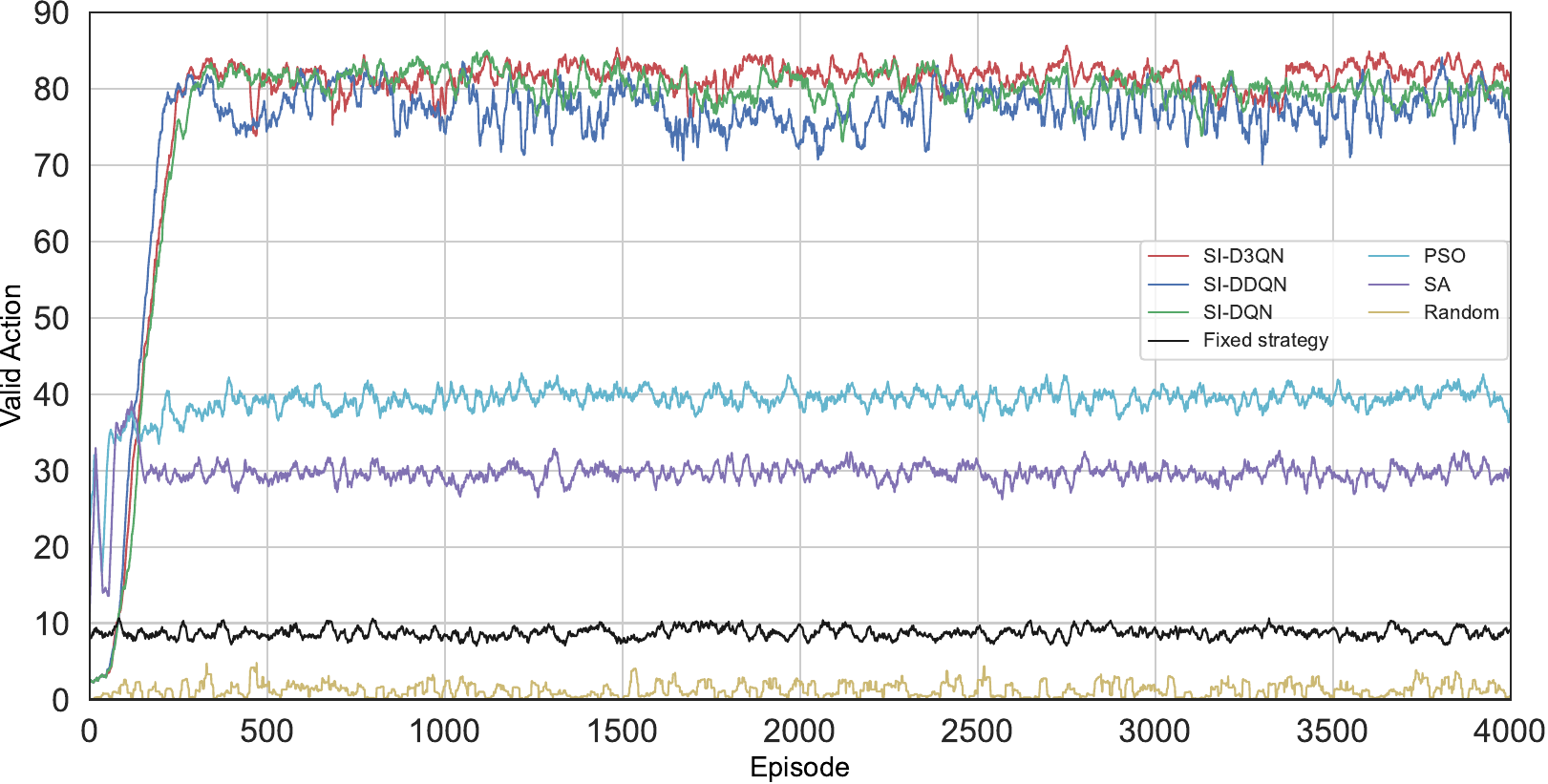}}
	\hspace{0.15in}
	\subfloat[]{\includegraphics[width=0.23\textwidth]{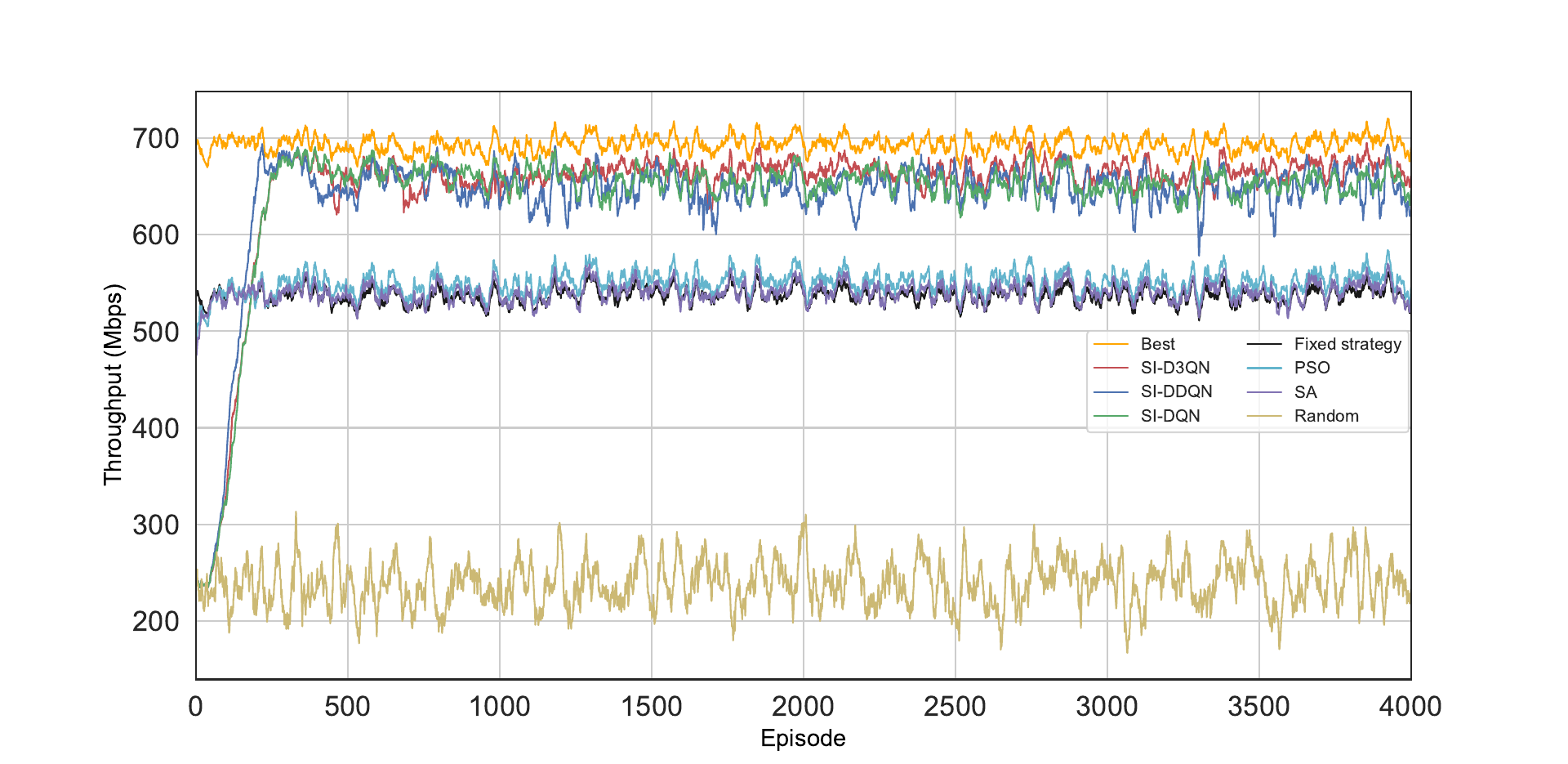}}
	\caption{The performance of SI-D3QN agent in Game 1. (a) Valid actions; (b) Throughput;}
	\label{fig4:env}
\end{figure}
\begin{figure}[ht]
	\subfloat[]{\includegraphics[width=0.23\textwidth]{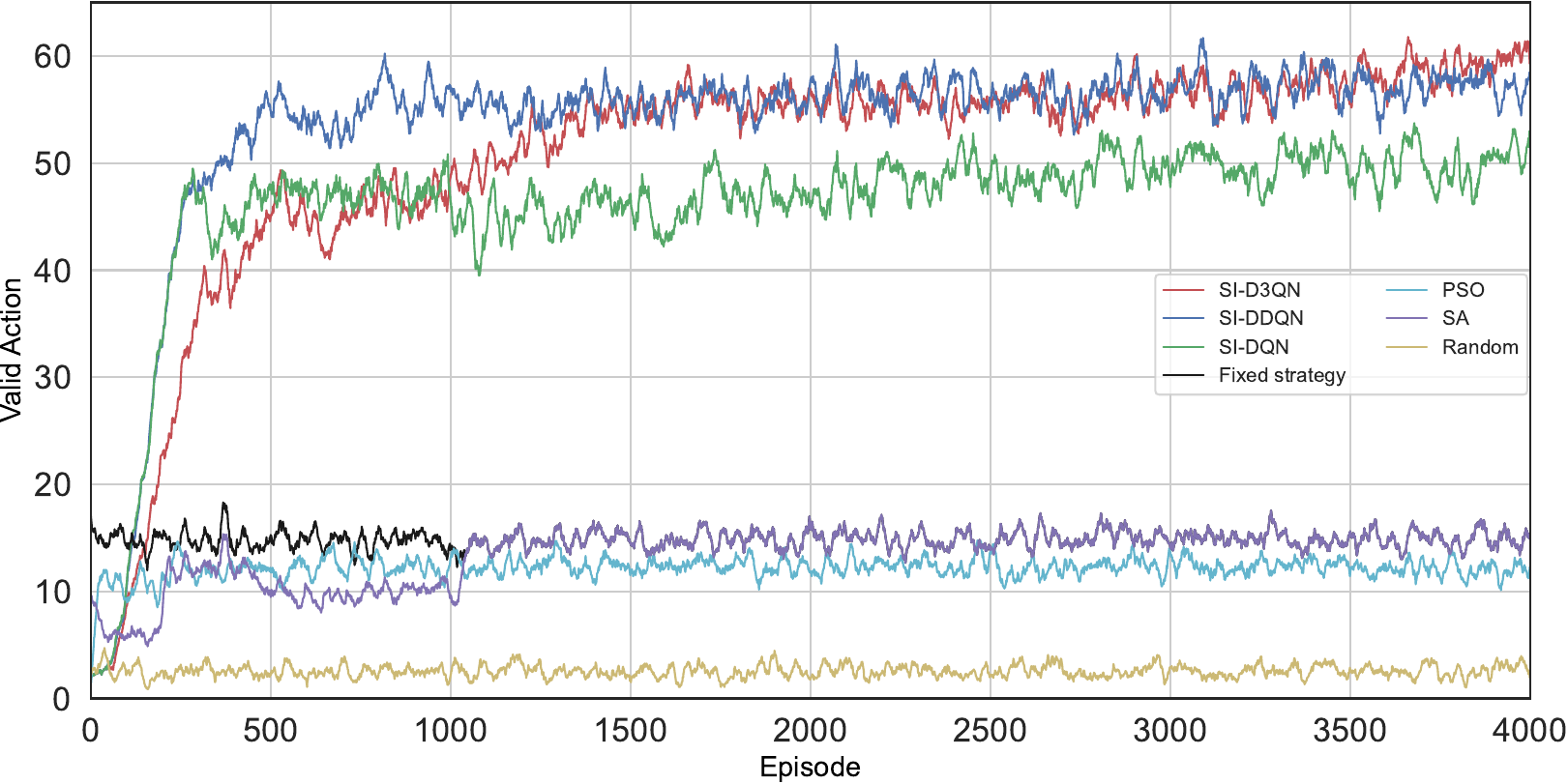}}
	\hspace{0.15in}
	\subfloat[]{\includegraphics[width=0.23\textwidth]{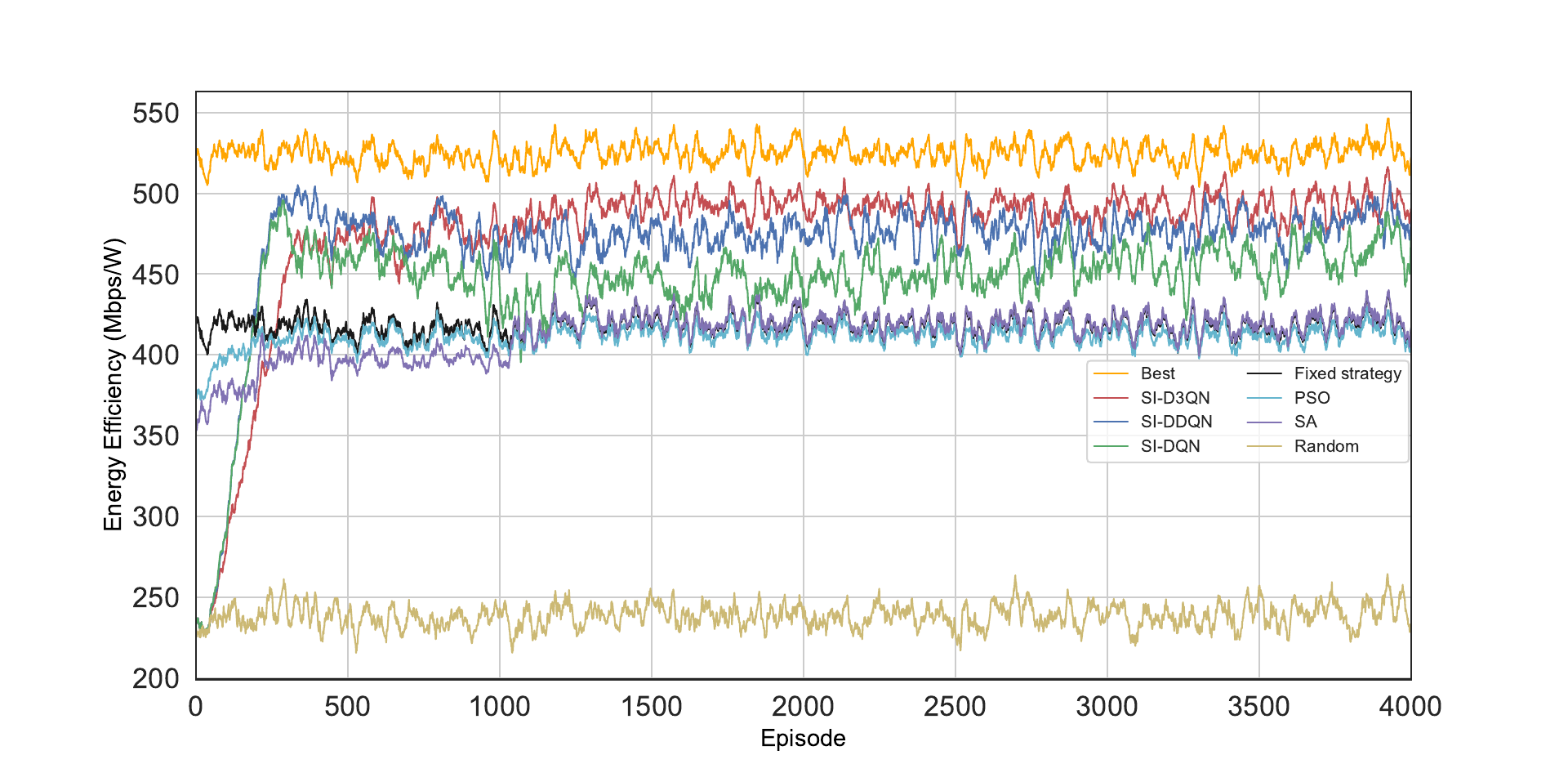}}
	\caption{The performance of SI-D3QN agent in Game 2. (a) Valid actions; (b) Energy efficiency;}
	\label{fig4:env}
\end{figure}
\subsection{Simulation Settings}\label{AA}
 The key  parameters and different algorithms simulation parameters are shown in Table I and Table II. Operating under the IEEE 802.11p, the V2V communication system is designed with a half-clocked mode, using a channel bandwidth of 10 MHz centered in the 5.9 GHz frequency spectrum. Additionally, it incorporates OFDM featuring 64 subcarriers in total as its transmission method. In the context of U-NLOS  vehicular scenario, we perform 100 transmissions for every eposide, where the channel conditions are randomly varied within the limited range of SNR for each transmission to simulate the complex vehicular communication environment under high dynamics. The action space for SI-D3QN agent in both Game 1 and Game 2 is 40, where these action set is aligned with industry standards.

\subsection{Results and Observations on Game 1 and Game 2}\label{AA}
The only difference between Game 1 and Game 2 lies in the optimization objective: Game 1 aims to optimize throughput, while Game 2 focuses on enhancing energy efficiency. Game 2 is more complex than Game 1, owing to the increased number of constraints under identical conditions, and the reward is entangled with these constraints. 

In order to better evaluate the performance of DRL agents, we compare different approaches and the best reward for each game. It is worth noting that the best reward is the reward upper limit obtained by the use of optimal policy and cannot be realized in practice. Besides, we assess the effectiveness of decisions through the dual indicators of the number of valid actions and the total accumulated rewards, where the decisions that maximize the reward during each transmission are referred to as valid actions.

The performance of different algorithms after 4000 training episodes in Game 1 and Game 2 is shown in Fig. 3 and Fig. 4 respectively. As seen from Fig. 3(a) and Fig. 4(a), the SI-D3QN agent maintains the highest number of effective decisions in the two games. Besides, from other part of the results, the agent's average energy efficiency in Game 1 is 496 Mbps/W and in Game 2 is 492.6 Mbps/W, while the average energy consumption to maintain the corresponding throughput in Game 1 and Game 2 is 1.04 W and 1.35 W. Therefore, it indicates that SI-D3QN agent obtains excellent long-term energy efficiency and has increased the link throughput by 29.6\% under the same energy consumption. This is because SI-D3QN algorithm reduces the overestimations and learns the optimal policy for all the state-action pairs more efficiently. Transmission design using traditional heuristic tools (such as SA and PSO) might not be feasible because the original optimization problem (9) is highly complex. However, the DRL agents can still  identify the solutions that are closer to the optimal.  It is evident that the SI-D3QN agent, across various tasks, is closer to optimal performance and achieves superior long-term rewards compared to other methods.

\section{Conclusion}
The letter focuses on the joint design of MCS selection and power control to enhance the energy efficiency in vehicular communications. Driven by SI technique, we present a novel DRL approach, named SI-D3QN. The algorithm employs two independent Dueling DQN networks for action-value evaluation and action selection separately, effectively mitigating the issue of overestimation during the training periods. Experiments in diverse scenarios and highly dynamic vehicular environments demonstrate that the SI-D3QN agent not only ensures effective decision-making but also significantly improves the long-term link perormance compared to DRL benchmark algorithms, and other traditional optimization algorithms, highlighting the advantages of the SI-D3QN agent for transmission design.

\bibliographystyle{ieeetr}
\bibliography{reference}
\vspace{12pt}
\color{red}
\end{sloppypar}
\end{document}